\documentstyle[iceccs]{article}
\newtheorem{definition}{Definition}[section]

\renewcommand{\baselinestretch}{.96}

\newcommand{\qqquad}{\quad\qquad}
\newcommand{\qqqquad}{\quad\qqquad}
\newcommand{\qqqqquad}{\quad\qqqquad}
\newcommand{\qqqqqquad}{\quad\qqqqquad}
\newcommand{\qqqqqqquad}{\quad\qqqqqquad}
\newcommand{\qqqqqqqquad}{\quad\qqqqqqquad}
\newcommand{\qqqqqqqqquad}{\quad\qqqqqqqquad}

\title{\Large\bf Applying Slicing Technique to Software Architectures}

\author{\large Jianjun Zhao\\
Department of Computer Science and Engineering\\
Fukuoka Institute of Technology\\
3-10-1 Wajiro-Higashi, Higashi-ku, Fukuoka 811-0214, Japan\\
Email: zhao@cs.fit.ac.jp}

\date{}
\begin{document}

\maketitle

\thispagestyle{empty}
{\normalsize
\abstract
{Software architecture is receiving increasingly attention as a
critical design level for software systems. As software architecture design resources (in the form of architectural specifications) are going to be 
accumulated, the development of techniques and tools to support
architectural understanding, testing, reengineering, maintenance, and reuse 
will become an important issue. This paper introduces a new form of slicing, 
named {\it architectural slicing}, to aid architectural understanding 
and reuse. In contrast to traditional slicing, architectural slicing is 
designed to operate on the architectural specification of 
a software system, rather than the source code of a program. 
Architectural slicing provides knowledge about the high-level
structure of a software system, rather than the low-level 
implementation details of a program. In order to compute an architectural slice, we present the {\it architecture 
information flow graph} which can be used to represent 
information flows in a software architecture. Based on the graph, we give 
a two-phase algorithm to compute an architectural slice.}

\vspace{3mm}
\section{Introduction}
\label{sec:intro}
\vspace{3mm}

Software architecture is receiving increasingly attention as a
critical design level for software systems \cite{Shaw96}. The software 
architecture of a system defines its high-level structure, exposing 
its gross organization as a collection of interacting components. 
A well-defined architecture allows an engineer to reason about system 
properties at a high level of abstraction. 
Architectural description languages (ADLs) are formal languages that
can be used to represent the architecture of a software system. 
They focus on the high-level structure of the overall application
rather than the implementation details of any specific source module. 
Recently, a number of architectural description languages
have been proposed such as W{\sc right}\cite{Allen97}, Rapide \cite{Luckham95}, UniCon \cite{Shaw95}, 
and ACME \cite{Garlan97} to support formal representation
and reasoning of software architectures. 
As software architecture design resources (in the form of
architectural specifications) are going to be accumulated, 
the development of techniques to support software architectural 
understanding, testing, reengineering, maintenance and reuse 
will become an important issue.  

One way to support software architecture development is to use 
slicing technique. Program slicing, originally introduced by Weiser 
\cite{Weiser79}, is a decomposition technique which extracts program elements 
related to a particular computation. A {\it program slice} consists 
of those parts of a program that may directly or indirectly affect the
values computed at some program point of interest, 
referred to as a {\it slicing criterion}. The task to compute program
slices is called {\it program slicing}. To understand the basic
idea of program slicing, consider a simple example in Figure
\ref{fig:Ada} which shows: (a) a program fragment and (b) its slice with
respect to the slice criterion (\verb+Total+,14). The slice consists 
of only those statements in the program that might affect the value of
variable \verb+Total+ at line 14. The lines represented by small rectangles 
are statements that have been sliced away. We refer to this kind of slicing 
as {\it traditional slicing} to distinguish it from a new form of slicing 
introduced later.

Traditional slicing has been studied primarily in the context of 
conventional programming languages \cite{Tip95}. In such languages, 
slicing is typically performed by using a control flow graph 
or a dependence graph 
\cite{Cheng93,Horwitz90,Ferrante87,Ottenstein84,Zhao96,Zhao97}. 
Traditional slicing has many applications in software engineering 
activities including program understanding \cite{DeLucia96}, 
debugging \cite{Agrawal93}, testing \cite{Bates93}, 
maintenance \cite{Gallagher91}, reuse \cite{Ning94}, 
reverse engineering \cite{Beck93}, and complexity measurement 
\cite{Ottenstein84}.

\begin{figure}[t]
  \begin{center}
     \epsfile{file=fig1.eps,scale=1}
  \caption{\label{fig:Ada} A program fragment and its slice
on criterion ({\protect\verb+Total+},14).}
  \end{center}
\end{figure}

Applying slicing technique to software architectures promises benefit 
for software architecture development at least in two aspects. 
First, architectural understanding and maintenance 
should benefit from slicing. When a maintainer wants to modify a
component in a software architecture in order to satisfy new design 
requirements, the maintainer must first investigate which components 
will affect the modified component and which components will be 
affected by the modified component. This process is usually called {\it 
impact analysis}. By slicing a software architecture, the maintainer 
can extract the parts of a software architecture containing 
those components that might affect, or be affected by, the modified 
component. The slicing tool which provides such information can assist 
the maintainer greatly. 
Second, architectural reuse should benefit from slicing. While reuse
of code is important, in order to make truly large gains in
productivity and quality, reuse of software designs and patterns may 
offer the greater potential for return on investment. 
By slicing a software architecture, a system designer can extract 
reusable architectures from it, and reuse them into new system designs 
for which they are appropriate.

While slicing is useful in software architecture development, 
existing slicing techniques for conventional programming languages 
can not be applied to architectural specifications straightforwardly 
due to the following reasons. Generally, the traditional definition of 
slicing is concerned with slicing programs written in conventional 
programming languages which primarily consist of variables and statements, 
and the slicing notions are usually defined as (1) a slicing criterion 
is a pair (s, V) where $s$ is a statement and $V$ is a set of variables defined
or used at $s$, and (2) a slice consists of only statements. However, 
in a software architecture, the basic elements are components and their 
interconnections, but neither variables nor statements as in 
conventional programming languages. Therefore, to perform slicing at 
the architectural level, new slicing notions for software architectures 
must be defined. 

In this paper, we introduce a new form of slicing, 
named {\it architectural slicing}. In contrast to 
traditional slicing, architectural slicing is designed to operate on 
a formal architectural specification of a software system, rather than
the source code of a conventional program. Architectural slicing provides 
knowledge about the high-level structure of a software system, 
rather than the low-level implementation details of a conventional program. 
Our purpose for development of architectural slicing is different 
from that for development of traditional slicing. While 
traditional slicing was designed originally for supporting source code
level understanding and debugging of conventional programs, 
architectural slicing was primarily designed for supporting
architectural level understanding and reuse of large-scale software systems. 
However, just as traditional slicing has many other 
applications in software engineering activities, we believe that
architectural slicing is also useful in other software architecture 
development activities including architectural testing, reverse engineering, 
reengineering, and complexity measurement. 

Abstractly, our slicing algorithm takes as input a formal architectural 
specification (written in its associated architectural description language) 
of a software system, then it removes from the specification those components 
and interconnections between components which are not necessary for 
ensuring that the semantics of the specification of the software architecture 
is maintained. This benefit allows unnecessary components and interconnections 
between components to be removed at the architectural level of 
the system which may lead to considerable space savings, especially 
for large-scale software systems whose architectures consist of numerous 
components. In order to compute an architectural slice, we present the {\it architecture 
information flow graph} which can be used to represent 
information flows in a software architecture. Based on the graph, we give 
a two-phase algorithm to compute an architectural slice.


The rest of the paper is organized as follows. 
Section \ref{sec:wright} briefly introduces how to represent a software architecture using W{\sc right}: an architectural description language. Section \ref{sec:example} shows 
a motivation example. Section \ref{sec:sslicing} defines some notions about
slicing software architectures. Section \ref{sec:depen} presents the 
architecture information flow graph for software architectures . 
Section \ref{sec:comp} gives a two-phase algorithm for computing an 
architectural slice. Section \ref{sec:work} discusses the related work. 
Concluding remarks are given in Section \ref{sec:final}.

\begin{figure*}[t]
  \renewcommand{\arraystretch}{0.85}
{\sf
\begin{center}
{\scriptsize
\begin{tabular}{l}
     {\bf Configuration} GasStation\\
     \qquad{\bf Component} Customer\\
     \qqqquad {\bf Port} Pay = \underline{pay!x} $\rightarrow$ Pay\\
     \qqqquad {\bf Port} Gas = \underline{take} $\rightarrow$ pump$?$x $\rightarrow$ Gas\\
     \qqqquad {\bf Computation} = \underline{Pay.pay!x} $\rightarrow$ \underline{Gas.take} $\rightarrow$ Gas.pump$?$x $\rightarrow$ Computation\\

     \qquad{\bf Component} Cashier\\
     \qqqquad {\bf Port} Customer1 = pay$?$x $\rightarrow$ Customer1\\
     \qqqquad {\bf Port} Customer2 = pay$?$x $\rightarrow$ Customer2\\
     \qqqquad {\bf Port} Topump = \underline{pump!x} $\rightarrow$ Topump\\
     \qqqquad {\bf Computation} = Customer1.pay$?$x $\rightarrow$ \underline{Topump.pump!x} $\rightarrow$ 
Computation\\
     \qqqqqquad [] Customer2.pay$?$x $\rightarrow$ \underline{Topump.pump!x} $\rightarrow$ Computation\\

     \qquad{\bf Component} Pump\\
     \qqqquad {\bf Port} Oil1 = take $\rightarrow$ \underline{pump!x} $\rightarrow$ Oil1\\
     \qqqquad {\bf Port} Oil2 = take $\rightarrow$ \underline{pump!x} $\rightarrow$ Oil2\\
     \qqqquad {\bf Port} Fromcashier = pump$?$x $\rightarrow$ Fromcashier\\
     \qqqquad {\bf Computation} = Fromcashier.pump$?$x $\rightarrow$ \\
     \qqqqqquad (Oil1.take $\rightarrow$ \underline{Oil1.pump!x} $\rightarrow$ Computation)\\
     \qqqqqquad [] (Oil2.take $\rightarrow$ \underline{Oil2.pump!x} $\rightarrow$ Computation)\\

     \qquad{\bf Connector} Customer\_Cashier\\
     \qqqquad {\bf Role} Givemoney = \underline{pay!x} $\rightarrow$ Givemoney\\
     \qqqquad {\bf Role} Getmoney = pay$?$x $\rightarrow$ Getmoney\\
     \qqqquad {\bf Glue} = Givemoney.pay$?$x $\rightarrow$ \underline{Getmoney.pay!x} $\rightarrow$ Glue\\

     \qquad{\bf Connector} Customer\_Pump\\
     \qqqquad {\bf Role} Getoil = \underline{take} $\rightarrow$ pump$?$x $\rightarrow$ Getoil\\
     \qqqquad {\bf Role} Giveoil = take $\rightarrow$ \underline{pump!x} $\rightarrow$ Giveoil\\
     \qqqquad {\bf Glue} = Getoil.take $\rightarrow$ \underline{Giveoil.take} $\rightarrow$ Giveoil.pump$?$x $\rightarrow$ \underline{Getoil.pump!x} $\rightarrow$ Glue\\

     \qquad{\bf Connector} Cashier\_Pump\\
     \qqqquad {\bf Role} Tell = \underline{pump!x} $\rightarrow$ Tell\\
     \qqqquad {\bf Role} Know = pump$?$x $\rightarrow$ Know\\
     \qqqquad {\bf Glue} = Tell.pump$?$x $\rightarrow$ \underline{Know.pump!x} $\rightarrow$ Glue\\

     \qquad{\bf Instances}\\
     \qqqquad Customer1: Customer\\
     \qqqquad Customer2: Customer\\
     \qqqquad cashier: Cashier\\
     \qqqquad pump: Pump\\
     \qqqquad Customer1\_cashier: Customer\_Cashier\\
     \qqqquad Customer2\_cashier: Customer\_Cashier\\
     \qqqquad Customer1\_pump: Customer\_Pump\\
     \qqqquad Customer2\_pump: Customer\_Pump\\
     \qqqquad cashier\_pump: Cashier\_Pump\\

     \qquad{\bf Attachments}\\
     \qqqquad Customer1.Pay as Customer1\_cashier.Givemoney\\
     \qqqquad Customer1.Gas as Customer1\_pump.Getoil\\
     \qqqquad Customer2.Pay as Customer2\_cashier.Givemoney\\
     \qqqquad Customer2.Gas as Customer2\_pump.Getoil\\
     \qqqquad casier.Customer1 as Customer1\_cashier.Getmoney\\
     \qqqquad casier.Customer2 as Customer2\_cashier.Getmoney\\
     \qqqquad cashier.Topump as cashier\_pump.Tell\\
     \qqqquad pump.Fromcashier as cashier\_pump.Know\\
     \qqqquad pump.Oil1 as Customer1\_pump.Giveoil\\
     \qqqquad pump.Oil2 as Customer2\_pump.Giveoil\\
     {\bf End} GasStation.\\
    \\
\end{tabular}
}
\end{center}
}

  \caption{\label{fig:wright} An architectural specification in W{\sc right}.}

\end{figure*}

\begin{figure}[t]
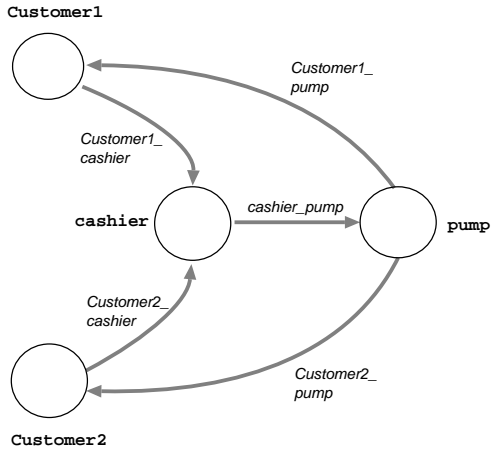

  \begin{center}
     \epsfile{file=fig2.eps,scale=0.85}
  \caption{\label{fig:sys} The architecture of the Gas Station system.}
  \end{center}
\end{figure}

\vspace{3mm}
\section{Software Architectural Specification in W{\sc right}}
\label{sec:wright}
\vspace{3mm}

We assume that readers are familiar with the basic concepts of software 
architecture and architectural description language, and in this paper, 
we use W{\sc right} architectural description language \cite{Allen97} as our target
language for formally representing software architectures. The selection of 
W{\sc right} is based on that it supports to represent not only the 
architectural structure but also the architectural behavior 
of a software architecture. 

Below, we use a simple W{\sc right} architectural specification taken 
from \cite{Naumovich97} as a sample to briefly introduce how 
to use W{\sc right} to represent a software architecture. The 
specification is showed in Figure \ref{fig:wright} which models 
the system architecture of a Gas Station system \cite{Helmbold85}. 

\vspace{2mm}
\subsection{Representing Architectural Structure}
\vspace{2mm}


W{\sc right} uses a {\it configuration} to describe architectural structure 
as graph of components and connectors. 

{\it Components} are computation units in the system. 
In W{\sc right}, each component has an {\it interface} 
defined by a set of {\it ports}. Each port identifies a point of 
interaction between the component and its environment. 

{\it Connectors} are patterns of interaction between components.
In W{\sc right}, each connector has an {\it interface} 
defined by a set of {\it roles}. Each role defines 
a participant of the interaction represented by the connector. 

A W{\sc right} architectural specification of a system is 
defined by a set of component and connector type definitions, 
a set of instantiations of specific objects of these types, 
and a set of {\it attachments}. Attachments specify which components 
are linked to which connectors.  
  
For example, in Figure \ref{fig:wright} there are three component 
type definitions, \verb+Customer+, \verb+Cashier+ and \verb+Pump+, 
and three connector type definitions, \verb+Customer_Cashier+, 
\verb+Customer_Pump+ and \verb+Cashier_Pump+. The configuration is composed 
of a set of instances and a set of attachments to specify the 
architectural structure of the system.

\vspace{2mm}
\subsection{Representing Architectural Behavior}
\vspace{2mm}

W{\sc right} models architectural behavior according to the significant 
events that take place in the computation of components, and the 
interactions between components as described by the connectors. 
The notation for specifying event-based behavior is adapted from 
CSP \cite{Hoare85}. Each CSP process defines an alphabet of events and the 
permitted patterns of events that the process may exhibit. These 
processes synchronize on common events (i.e., interact) 
when composed in parallel. W{\sc right} uses such process descriptions 
to describe the behavior of ports, roles, computations and glues.

A {\it computation} specification specifies a component's behavior: 
the way in which it accepts certain events on certain {\it ports} 
and produces new events on those or other ports. Moreover, W{\sc right} 
uses an overbar to distinguish initiated events from observed events
\footnote{In this paper, we use an underbar to represent an initiated 
event instead of an overbar that used in the original W{\sc right} 
language definition \cite{Allen97}.}. 
For example, the \verb+Customer+ 
initiates \verb+Pay+ action (i.e., \underline{\verb+pay!x+}) 
while the \verb+Cashier+ observes it (i.e., \verb+pay?x+).
  
A {\it port} specification specifies the local protocol with which the 
component interacts with its environment through that port. 

A {\it role} specification specifies the protocol that must be satisfied 
by any port that is attached to that role. 
Generally, a port need no have the same behavior as the role that it fills, 
but may choose to use only a subset of the connector capabilities. 
For example, the \verb+Customer+ role \verb+Gas+ and the \verb+Customer_Pump+ 
port \verb+Getoil+ are identical.

A {\it glue} specification specifies how the roles of a connector interact 
with each other. For example, a \verb+Cashier_Pump+ tell (Tell.pump$?$x) must 
be transmitted to the \verb+Cashier_Pump+ know (\underline{\verb+Know.pump!x+}).

As a result, based on formal W{\sc right} architectural specifications, 
we can infer which ports of a component are input ports and which are 
output ports. Also, we can infer which roles are input roles and which are 
output roles. Moreover, the direction in which the information transfers 
between ports and/or roles can also be inferred based on the 
formal specification. As we will show in Section \ref{sec:depen}, such 
kinds of information can be used to construct the information flow graph 
for a software architecture for computing an architectural slice 
efficiently.
 
\vspace{2mm}
In order to focus on the key ideas of architectural slicing, in this paper 
we assume that a software architecture be represented by a formal 
architectural specification which contains three basic types of design 
entities, namely, {\it components} whose interfaces are defined by a set 
of elements called {\it ports}, {\it connectors} whose interfaces are 
defined by a set of elements called {\it roles} and 
the {\it configuration} whose topology is declared by a set of elements 
called {\it instances} and {\it attachments}. Moreover, each component has a 
special element called {\it computation} and each connector has a special 
element called {\it glue} as we described above.

In the rest of the paper, we assume that an architectural specification $P$ 
be denoted by $(C_{m}, C_{n}, c_{g})$ where: 
\begin{itemize}
	\item $C_{m}$ is the set of components in $P$,
	\item $C_{n}$ is the set of connectors in $P$, and 
	\item $c_{g}$ is the configuration of $P$.
\end{itemize}

\vspace{3mm}
\section{Motivation Example}
\label{sec:example}
\vspace{3mm}

We present a simple example to explain our approach on 
architectural slicing. The example also shows one application 
of architectural slicing, in which it is used in the impact analysis 
of software architectures.

Consider the Gas Station system whose architectural representation 
is shown in Figure \ref{fig:sys}, and W{\sc right} specification is shown in 
Figure \ref{fig:wright}. Suppose a maintainer needs to modify the 
component \verb+cashier+ in the architectural specification in 
order to satisfy some new design requirements. The first thing 
the maintainer has to do is to investigate which components 
and connectors interact with component \verb+cashier+ through 
its ports \verb+Customer1+, \verb+Customer2+, and \verb+Topump+. 
A common way is to manually check 
the source code of the specification to find such information. However, 
it is very time-consuming and error-prone even for a small size 
specification because there may be complex dependence relations 
between components in the specification. If the maintainer has an architectural slicer at hand, the work may probably be simplified and automated without the 
disadvantages mentioned above. In such a scenario, an architectural slicer 
is invoked, which takes as input: (1) a complete architectural specification 
of the system, and (2) a set of ports of the component \verb+cashier+, 
i.e., \verb+Customer1+, \verb+Customer2+ and \verb+Topump+ (this is an 
{\it architectural slicing criterion}). The slicer then computes a 
backward 
and forward architectural slice respectively with respect to the 
criterion and outputs them to the maintainer. 
A backward architectural slice is a partial specification of the original one 
which includes those components and connectors that might affect the 
component \verb+cashier+ through the ports in the criterion, and a forward 
architectural slice is a partial specification of the original 
one which includes those components and connectors that might be
affected by the component \verb+cashier+ through the ports in the
criterion. The other parts of the specification that might not affect or
be affected by the component \verb+cashier+ will be removed, 
i.e., sliced away from the original specification. The maintainer can 
thus examine only the contents included in a slice to investigate 
the impact of modification. Using the algorithm we will present in 
Section \ref{sec:comp}, the slice shown in Figure \ref{fig:wright.slice} 
can be computed.

\vspace{3mm}
\section{Architectural Slicing}
\label{sec:sslicing}
\vspace{3mm}

Intuitively, an {\it architectural slice} may be viewed as a subset 
of the behavior of a software architecture, similar to the original notion 
of the traditional static slice. However, while a traditional slice intends 
to isolate the behavior of a specified set of program variables, 
an architectural slice intends to isolate the behavior of a specified set 
of a component or connector's elements. Given an architectural 
specification $P=(C_{m}, C_{n}, c_{g})$, our goal is to compute an 
architectural slice $S_{p}=(C_{m}', C_{n}', c_{g}')$ which should 
be a ``sub-architecture'' of $P$ and preserve partially the 
semantics of $P$. To define the meanings of the word ``sub-architecture,'' 
we introduce the concepts of a reduced component, connector and 
configuration. 

\begin{definition}\label{def:rcomp}
	Let $P=(C_{m}, C_{n}, c_{g})$ be an architectural specification 
        and $c_{m} \in C_{m}$, $c_{n} \in C_{n}$, and $c_{g}$ 
        be a component, connector, and configuration of $P$ 
        respectively: 

	\begin{itemize} 
		\item A {\em reduced component} of $c_{m}$ is a
		      component $c_{m}'$ that is derived from
                      $c_{m}$ by removing zero, or more elements from $c_{m}$. 

		\item A {\em reduced connector} of $c_{n}$ is a
		      connector $c_{n}'$ that is derived from
		      $c_{n}$ by removing zero, or more elements from $c_{n}$.

 		\item A {\em reduced configuration} of $c_{g}$ is a 
                      configuration $c_{g}'$ that is derived 
                      from $c_{g}$ by removing zero, or more elements 
                      from $c_{g}$. 
	\end{itemize} 
\end{definition}

The above definition showed that a reduced component, connector, 
or configuration of a component, connector, or configuration may equal 
itself in the case that none of its elements has been removed, 
or an {\it empty} component, connector, or 
configuration in the case that all its elements have been removed.

For example, the followings show a component \verb+Customer+, a
connector \verb+Customer_Cashier+, and a configuration as well as 
their corresponding reduced component, connector, and configuration. 
The small rectangles represent those ports, roles, or instances 
and attachments that have been removed from the original component, 
connector, or configuration.\\
 
(1) The component \verb+Customer+ and its reduced component (with * mark) 
    in which the port \verb+Gas+ and elements \underline{\verb+Gas.take+} and 
    \verb+Gas.pump?x+ that are related to \verb+Gas+ in the computation have 
 been removed.

{\sf\footnotesize
\begin{center}
\begin{tabular}{l}
     {\bf Component} Customer\\
     \qquad {\bf Port} Pay = \underline{pay!x} $\rightarrow$ Pay\\
     \qquad {\bf Port} Gas = \underline{take} $\rightarrow$ pump$?$x $\rightarrow$ Gas\\
     \qquad {\bf Computation} = \underline{Pay.pay!x} $\rightarrow$ \underline{Gas.take}  \\
     \qqqqqqqqquad $\rightarrow$ Gas.pump$?$x $\rightarrow$ Computation\\
     \\

     {\bf * Component} Customer\\
     \qquad {\bf Port} Pay = \underline{pay!x} $\rightarrow$ Pay\\
     \qquad $\Box$$\Box$$\Box$$\Box$$\Box$$\Box$$\Box$$\Box$$\Box$$\Box$$\Box$$\Box$$\Box$$\Box$$\Box$$\Box$$\Box$$\Box$$\Box$$\Box$\\
     \qquad {\bf Computation} = \underline{Pay.pay!x} $\rightarrow$ $\Box$$\Box$$\Box$$\Box$$\Box$$\Box$  \\
     \qqqqqqqqquad $\rightarrow$ $\Box$$\Box$$\Box$$\Box$$\Box$$\Box$ $\rightarrow$ Computation\\
\end{tabular}
\end{center}
}

(2) The connector \verb+Customer_Cashier+ and its reduced connector 
    (with * mark) in which the role \verb+Givemoney+ and the element 
    \verb+Givemoney.pay?x+ that is related to \verb+Givemoney+ in the glue 
    have been removed.

{\sf\footnotesize
\begin{center}
\begin{tabular}{l}
     {\bf Connector} Customer\_Cashier\\
     \qquad {\bf Role} Givemoney = \underline{pay!x} $\rightarrow$ Givemoney\\
     \qquad {\bf Role} Getmoney = pay$?$x $\rightarrow$ Getmoney\\
     \qquad {\bf Glue} = Givemoney.pay$?$x $\rightarrow$ \underline{Getmoney.pay!x} \\
     \qqqqqquad $\rightarrow$ Glue\\
      \\
     {\bf * Connector} Customer\_Cashier\\
     \qquad $\Box$$\Box$$\Box$$\Box$$\Box$$\Box$$\Box$$\Box$$\Box$$\Box$$\Box$$\Box$$\Box$$\Box$$\Box$$\Box$$\Box$$\Box$$\Box$$\Box$\\
     \qquad {\bf Role} Getmoney = pay$?$x $\rightarrow$ Getmoney\\
     \qquad {\bf Glue} = $\Box$$\Box$$\Box$$\Box$$\Box$$\Box$$\Box$$\Box$$\Box$$\Box$ $\rightarrow$ \underline{Getmoney.pay!x}\\
     \qqqqqquad $\rightarrow$ Glue\\

\end{tabular}
\end{center}
}

(3) The configuration and its reduced configuration (with * mark) in which 
    some instances and attachments have been removed.

{\renewcommand{\arraystretch}{0.9}
{\sf
\begin{center}
{\footnotesize
\begin{tabular}{l}
     {\bf Instances}\\
     \qquad Customer1: Customer\\
     \qquad Customer2: Customer\\
     \qquad cashier: Cashier\\
     \qquad pump: Pump\\
     \qquad Customer1\_cashier: Customer\_Cashier\\
     \qquad Customer2\_cashier: Customer\_Cashier\\
     \qquad Customer1\_pump: Customer\_Pump\\
     \qquad Customer2\_pump: Customer\_Pump\\
     \qquad cashier\_pump: Cashier\_Pump\\
     
     {\bf Attachments}\\
     \qquad Customer1.Pay as Customer1\_cashier.Givemoney\\
     \qquad Customer1.Gas as Customer1\_pump.Getoil\\
     \qquad Customer2.Pay as Customer2\_cashier.Givemoney\\
     \qquad Customer2.Gas as Customer2\_pump.Getoil\\
     \qquad casier.Customer1 as Customer1\_cashier.Getmoney\\
     \qquad casier.Customer2 as Customer2\_cashier.Getmoney\\
     \qquad cashier.Topump as cashier\_pump.Tell\\
     \qquad pump.Fromcashier as cashier\_pump.Know\\
     \qquad pump.Oil1 as Customer1\_pump.Giveoil\\
     \qquad pump.Oil2 as Customer2\_pump.Giveoil\\
    \\
\end{tabular}
}
\end{center}
}

{\renewcommand{\arraystretch}{0.9}
{\sf
\begin{center}
{\footnotesize
\begin{tabular}{l}
     {\bf * Instances}\\
     \qquad Customer1: Customer\\
     \qquad Customer2: Customer\\
     \qquad cashier: Cashier\\
     \qquad $\Box$$\Box$$\Box$$\Box$$\Box$$\Box$$\Box$$\Box$\\
     \qquad Customer1\_cashier: Customer\_Cashier\\
     \qquad Customer2\_cashier: Customer\_Cashier\\
     \qquad $\Box$$\Box$$\Box$$\Box$$\Box$$\Box$$\Box$$\Box$$\Box$$\Box$$\Box$$\Box$$\Box$$\Box$$\Box$$\Box$$\Box$$\Box$$\Box$$\Box$$\Box$\\
     \qquad $\Box$$\Box$$\Box$$\Box$$\Box$$\Box$$\Box$$\Box$$\Box$$\Box$$\Box$$\Box$$\Box$$\Box$$\Box$$\Box$$\Box$$\Box$$\Box$$\Box$$\Box$\\
     \qquad $\Box$$\Box$$\Box$$\Box$$\Box$$\Box$$\Box$$\Box$$\Box$$\Box$$\Box$$\Box$$\Box$$\Box$$\Box$$\Box$$\Box$$\Box$\\
     {\bf *Attachments}\\
     \qquad Customer1.Pay as Customer1\_cashier.Givemoney\\
\qquad $\Box$$\Box$$\Box$$\Box$$\Box$$\Box$$\Box$$\Box$$\Box$$\Box$$\Box$$\Box$$\Box$$\Box$$\Box$$\Box$$\Box$$\Box$$\Box$$\Box$$\Box$$\Box$$\Box$$\Box$$\Box$$\Box$$\Box$\\
     \qquad Customer2.Pay as Customer2\_cashier.Givemoney\\
\qquad $\Box$$\Box$$\Box$$\Box$$\Box$$\Box$$\Box$$\Box$$\Box$$\Box$$\Box$$\Box$$\Box$$\Box$$\Box$$\Box$$\Box$$\Box$$\Box$$\Box$$\Box$$\Box$$\Box$$\Box$$\Box$$\Box$$\Box$\\
     \qquad casier.Customer1 as Customer1\_cashier.Getmoney\\
     \qquad casier.Customer2 as Customer2\_cashier.Getmoney\\
     \qquad $\Box$$\Box$$\Box$$\Box$$\Box$$\Box$$\Box$$\Box$$\Box$$\Box$$\Box$$\Box$$\Box$$\Box$$\Box$$\Box$$\Box$$\Box$$\Box$$\Box$$\Box$$\Box$$\Box$$\Box$\\
     \qquad $\Box$$\Box$$\Box$$\Box$$\Box$$\Box$$\Box$$\Box$$\Box$$\Box$$\Box$$\Box$$\Box$$\Box$$\Box$$\Box$$\Box$$\Box$$\Box$$\Box$$\Box$$\Box$$\Box$$\Box$$\Box$$\Box$$\Box$\\
     \qquad $\Box$$\Box$$\Box$$\Box$$\Box$$\Box$$\Box$$\Box$$\Box$$\Box$$\Box$$\Box$$\Box$$\Box$$\Box$$\Box$$\Box$$\Box$$\Box$$\Box$$\Box$$\Box$$\Box$$\Box$\\
     \qquad $\Box$$\Box$$\Box$$\Box$$\Box$$\Box$$\Box$$\Box$$\Box$$\Box$$\Box$$\Box$$\Box$$\Box$$\Box$$\Box$$\Box$$\Box$$\Box$$\Box$$\Box$$\Box$$\Box$$\Box$\\
    \\

\end{tabular}
}
\end{center}
}


Having the definitions of a reduced component, connector and 
configuration, we can define the meaning of the word ``sub-architecture''.

\begin{definition}\label{def:rad}
	Let $P=(C_{m}, C_{n}, c_{g})$ and $P'=(C_{m}', C_{n}',
c_{g}')$ be two architectural specifications.  
Then $P'$ is a {\em reduced architectural specification} of $P$ 
if: 

	\begin{itemize} 
		\item $C_{m}'=\{c_{m_{1}}',c_{m_{2}}',\ldots,
c_{m_{k}}'\}$ is a ``subset'' of $C_{m}=\{c_{m_{1}},c_{m_{2}},\ldots, 
c_{m_{k}}\}$ such that for $i=1,2,\ldots,k$, $c_{m_{i}}'$ is a 
reduced component of $c_{m_{i}}$, 

		\item $C_{n}'=\{c_{n_{1}}',c_{n_{2}}',\ldots,
c_{n_{k}}'\}$ is a ``subset'' of
$C_{n}=\{c_{n_{1}},c_{n_{2}},\ldots,c_{n_{k}}\}$ such that for 
$i=1,2,\ldots,k$, $c_{n_{i}}'$ is a reduced connector of $c_{n_{i}}$, 

		\item $c_{g}'$ is a reduced configuration of $c_{g}$, 
	\end{itemize} 
\end{definition}

Having the definition of a reduced architectural specification, 
we can define some notions about slicing software architectures.

In a W{\sc right} architectural specification, for example, 
a component's interface is defined to be 
a set of ports which identify the form of the component interacting with 
its environment, and a connector's interface is defined to be a set of roles 
which identify the form of the connector interacting with its environment. 
To understand how a component interacts with 
other components and connectors for making changes, a maintainer must 
examine each port of the component of interest. Moreover, it has been 
frequently emphasized that connectors are as important as components for 
architectural design, and a maintainer may also want to modify a connector 
during the maintenance. To satisfy these requirements, for example, 
we can define a slicing criterion for a W{\sc right} architectural 
specification as a set of ports of a component or a set of roles of a 
connector of interest.

\begin{definition}
Let $P=(C_{m}, C_{n}, c_{g})$ be an architectural specification.
A {\em slicing criterion} for $P$ is a pair $(c, E)$ such that:

\begin{itemize}
	\item[1.]  $c \in C_{m}$ and $E$ is a set of elements of $c$, or  
	\item[2.]  $c \in C_{n}$ and $E$ is a set of elements of $c$.
\end{itemize}
\end{definition}

Note that the selection of a slicing criterion depends on users' interests 
on what they want to examine. If they are interested in examining a 
component in an architectural specification, they may use slicing criterion 1. 
If they are interested in examining a connector, they may use slicing 
criterion 2.  Moreover, the determination of the set $E$ also depends 
on users' interests on what they want to examine. If they want to examine a 
component, then $E$ may be the set of ports or just a subset of ports 
of the component. If they want to examine a connector, 
then $E$ may be the set of roles or just a subset of roles of the connector. 

\begin{definition}\label{def:slice}
	Let $P=(C_{m}, C_{n}, c_{g})$ be an architectural specification. 
\begin{itemize}
	\item A {\em backward architectural slice} 
              $S_{bp}=(C_{m}', C_{n}', C_{g}')$ of $P$ on a given slicing 
              criterion $(c, E)$ is a reduced architectural specification 
              of $P$ which contains only those reduced components, connectors, 
              and configuration that might directly or indirectly affect 
              the behavior of $c$ through elements in $E$. 
	\item {\em Backward-slicing} an architectural specification $P$ on 
               a given slicing criterion is to find the backward 
               architectural slice of $P$ with respect to the criterion. 
\end{itemize}
\end{definition}

\begin{definition}\label{def:fslice}
	Let $P=(C_{m}, C_{n}, c_{g})$ be an architectural specification. 
\begin{itemize}
	\item A {\em forward architectural slice} 
              $S_{fp}=(C_{m}', C_{n}', C_{g}')$ of $P$ on a given slicing 
              criterion $(c, E)$ is a reduced architectural specification 
              of $P$ which contains only those reduced components, 
              connectors, and configuration that might be directly or 
              indirectly affected by the behavior of $c$ through elements 
              in $E$. 
	\item {\em Forward-slicing} an architectural specification $P$ on 
               a given slicing criterion is to find the forward architectural 
               slice of $P$ with respect to the criterion. 
\end{itemize}
\end{definition}

From Definitions \ref{def:slice} and \ref{def:fslice}, it is obviously
that there is at least one backward slice and at least one forward slice 
of an architectural specification that is the specification itself. Moreover, 
the architecture represented by $S_{bp}$ or $S_{fp}$ should be 
a ``sub-architecture'' of the architecture represented by $P$.

Defining an architectural slice as a reduced architectural specification 
of the original one is particularly useful for supporting 
architectural reuse. By using an architectural slicer, a system designer
can automatically decompose an existing architecture (in the case that
its architectural specification is available) into some small 
architectures each having its own functionality which may be reused 
in new system designs. Moreover, the view of an architectural slice 
as a reduced architectural specification dose not reduce its usefulness 
when applied it to architectural understanding because it also contains enough
information for a maintainer to facilitate the modification.

\begin{figure*}[t]
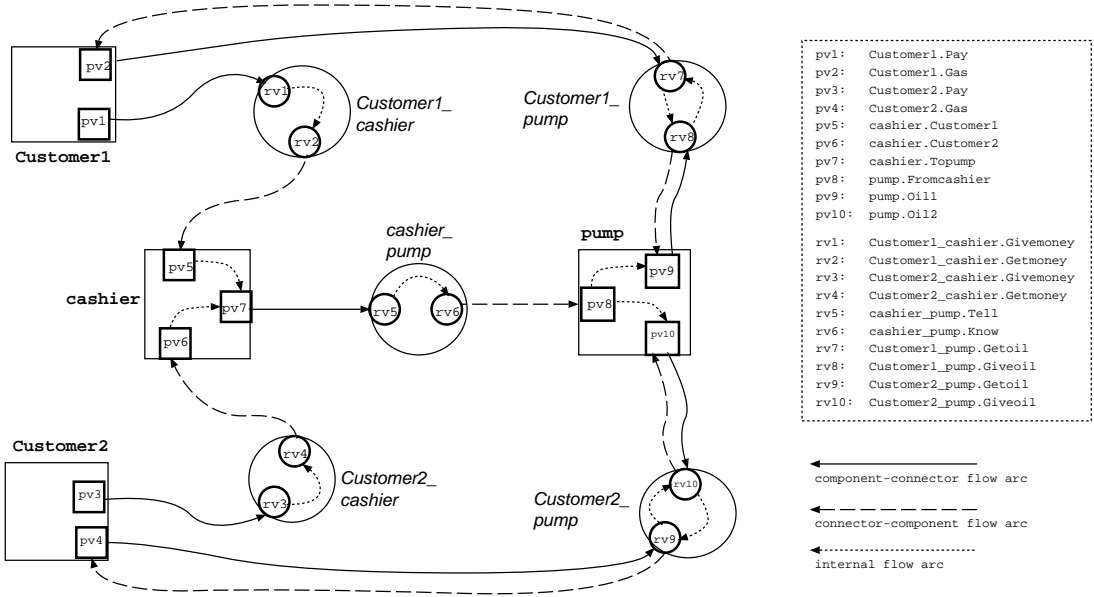

  \begin{center}
     \epsfile{file=fig3.eps,scale=0.85}
  \caption{\label{fig:graph} The information flow graph of the architectural
specification in Figure {\protect\ref{fig:wright}}.}
  \end{center}
\end{figure*}

\vspace{3mm}
\section{The Information Flow Graph for Software Architectures}
\label{sec:depen}
\vspace{3mm}

In this section, we present the architecture information flow graph 
for software architectures on which architectural slices can be 
computed efficiently. 

The architecture information flow graph is an arc-classified digraph 
whose vertices represent the ports of 
components and the roles of the connectors in an architectural specification, 
and arcs represent possible information flows between components and/or 
connectors in the specification.

\begin{definition}
	The {\em Architecture Information Flow Graph}\ (AIFG) of an 
        architectural specification $P$ is an arc-classified digraph 
        $(V_{com}, V_{con}, Com, Con, Int)$, where: 
	\begin{itemize}
		\item $V_{com}$ is the set of port vertices of $P$;
		\item $V_{con}$ is the set of role vertices of $P$;
		\item $Com$ is the set of component-connector flow arcs; 
		\item $Con$ is the set of connector-component flow arcs;
		\item $Int$ is the set of internal flow arcs.
	\end{itemize}
\end{definition}

There are three types of information flow arcs in the AIFG, namely, 
{\it component-connector flow arcs}, {\it connector-component flow arcs}, 
and {\it internal flow arcs}.

Component-connector flow arcs are used to represent information flows 
between a port of a component and a role of a connector in an architectural 
specification. Informally, if there is an information flow from a port 
of a component to a role of a connector in the specification, then there 
is a component-connector flow arc in the AIFG which connects the 
corresponding port vertex to the corresponding role vertex. 
For example, from the W{\sc right} specification shown in Figure \ref{fig:wright}, 
we can know that there is an information flow from the 
port \verb+Topump+ of the component \verb+cashier+ to 
the role \verb+Tell+ of the connector \verb+cashier_pump+. 
Therefore there is a component-connector flow arc in the AIFG in 
Figure \ref{fig:graph} which connects the port vertex of port 
\verb+Topump+ to the role vertex of role \verb+Tell+.

Connector-component flow arcs are used to represent information flows 
between a role of a connector and a port of a component in an architectural 
specification. Informally, if there is an information flow from a role of 
a connector to a port of a component in the specification, then there 
is a connector-component flow arc in the AIFG which connects the 
corresponding role vertex to the corresponding port vertex. 
For example, 
from the W{\sc right} specification in Figure \ref{fig:wright}, we can know that there 
is an information flow from the role \verb+Know+ of the connector 
\verb+cashier_pump+ to the port \verb+Fromcashier+ of the 
component \verb+pump+. Therefore, there is a connector-component 
flow arc in the AIFG in Figure \ref{fig:graph} which connects the role vertex 
for role \verb+Know+ to the port vertex for port \verb+Fromcashier+.

Internal flow arcs are used to represent internal information flows 
within a component or connector in an architectural specification. 
Informally, for a component in the specification, there is an internal 
flow from an input port to an output port, and for a connector 
in the specification, there is an internal flow from an input role 
to an output role. 
For example, in Figure \ref{fig:wright}, there is an internal flow 
from the role \verb+Givemoney+ to the role \verb+Getmoney+ of 
the connector \verb+Customer1_cashier+ and also an internal flow arc 
from the port \verb+Fromcashier+ to the port \verb+Oil1+ of component \verb+pump+.

As we introduced in Section \ref{sec:wright}, W{\sc right} uses CSP-based 
model to specify the behavior of a component and a connector of a 
software architecture. W{\sc right} allows user to infer which ports of 
a component are input and which are output, and which roles of a 
connector are input and which are output based on a W{\sc right} 
architectural specification. Moreover, it also allows user to infer 
the direction in which the information transfers between ports and/or roles. 
As a result, by using a static analysis tool which takes an architectural 
specification as its input, we can construct the AIFG of a 
W{\sc right} architectural specification  automatically.

Figure \ref{fig:graph} shows the AIFG of 
the architectural specification in Figure \ref{fig:wright}. In the figure, 
large squares represent components in the specification, and small
squares represent the ports of each component. Each port vertex has 
a name described by {\it component\_name.port\_name}. For example, 
$pv5$ (\verb+cashier.Customer1+) is a port vertex 
that represents the port \verb+Customer1+ of the 
component \verb+cashier+. 
Large circles represent connectors in the specification, 
and small circles represent the roles of each connector.  Each role
vertex has a name described by {\it connector\_name.role\_name}. 
For example, $rv5$ (\verb+cashier_pump.Tell+) 
is a role vertex that represents the role 
\verb+Tell+ of the connector \verb+cashier_pump+. The
complete specification of each vertex is shown on the right side 
of the figure.

Solid arcs represent component-connector flow arcs that 
connect a port of a component to a role of a connector. Dashed arcs represent 
connector-component flow arcs that connect a role of a connector to a port 
of a component. Dotted arcs represent internal flow arcs that connect 
two ports within a component (from an input port to an output port), 
or two roles 
within a connector (from an input role to an output role). 
For example, $(rv2,pv5)$ and $(rv6,pv8)$ are connector-component flow arcs. 
$(pv7,rv5)$ and $(pv9,rv8)$ are component-connector flow arcs. $(rv1,rv2)$ 
and $(pv8,pv10)$ are internal flow arcs.
 
\begin{figure*}[t]
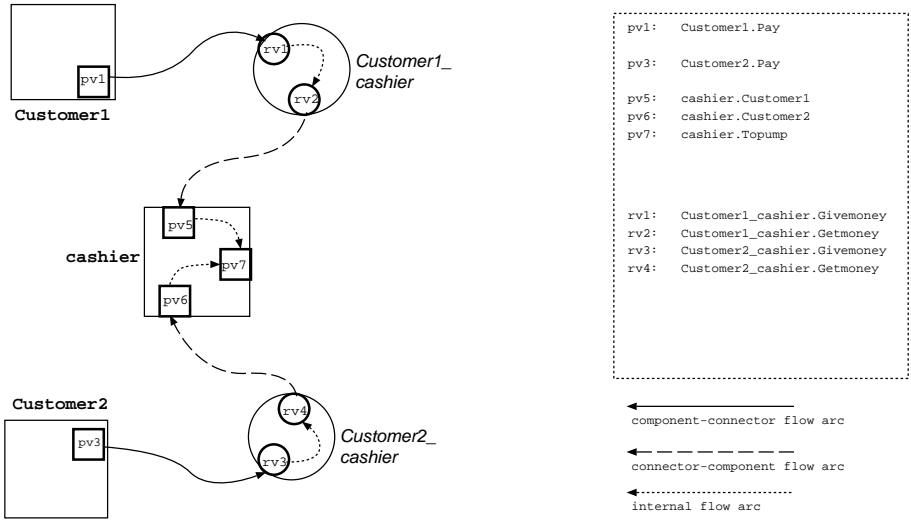

  \begin{center}
     \epsfile{file=fig4.eps,scale=0.85}
  \caption{\label{fig:graph.slice} A slice over the AIFG of the 
architectural specification in Figure {\protect\ref{fig:wright}}.}
  \end{center}
\end{figure*}

\vspace{3mm}
\section{Computing Architectural Slices}
\label{sec:comp}
\vspace{3mm}

The slicing notions defined in Section \ref{sec:sslicing} give us 
only a general view
of an architectural slice, and do not tell us how to compute it. 
In this section we present a two-phase algorithm to compute a slice of an
architectural specification based on its information flow graph. 
Our algorithm contains two phases: (1) Computing a slice $S_{g}$ over the 
information flow graph of an architectural specification, 
and (2) Constructing an architectural slice $S_{p}$ from $S_{g}$.

\vspace{2mm}
\subsection{Computing a Slice over the AIFG}
\vspace{2mm}

Let $P=(C_{m}, C_{n}, c_{g})$ be an architectural specification 
and $G=(V_{com},V_{con}, Com, Con, Int)$ be the AIFG of $P$. 
To compute a slice 
over the $G$, we refine the slicing notions defined in 
Section \ref{sec:sslicing} as follows: 

{\it 
\begin{itemize} 
	\item A {\em slicing criterion} for $G$ is a pair $(c,V_{c})$
	      such that: (1) $c \in C_{m}$ and $V_{c}$ is a set of port 
              vertices corresponding to the ports of $c$, or (2) $c
              \in C_{n}$ and $V_{c}$ is a set of role vertices
              corresponding to roles of $c$.

	\item The {\em backward slice} $S_{bg}(c,V_{c})$ of $G$ on a given 
              slicing
	      criterion $(c,V_{c})$ is a subset of vertices of $G$ such
	      that for any vertex $v$ of $G$, $v \in S_{bg}(c,V_{c})$ iff 
	      there exists a path from $v$ to $v' \in V_{c}$ in the AIFG.

	\item The {\em forward slice} $S_{fg}(c,V_{c})$ of $G$ on a
	      given slicing criterion $(c,V_{c})$ is a subset of
              vertices of $G$ such that for any vertex $v$ of $G$, $v
              \in S_{fg}(c,V_{c})$ iff there exists a path from 
              $v' \in V_{c}$ to $v$ in the AIFG.
\end{itemize} 
}

According to the above descriptions, the computation of a backward slice 
or forward slice over the AIFG can be solved by using an usual
depth-first or breath-first graph traversal algorithm to traverse the 
graph by taking some port or role vertices of interest as the start 
point of interest. 

Figure \ref{fig:graph.slice} shows a backward slice over the AIFG with respect
to the slicing criterion $(\verb+cashier+, V_{c})$ such that 
$V_{c}=\{pv5, pv6, pv7\}$. 

\vspace{2mm}
\subsection{Computing an Architectural Slice}
\vspace{2mm}

The slice $S_{g}$ computed above is only a slice
over the AIFG of an architectural specification, which is a set of
vertices of the AIFG. Therefore we should map each element in
$S_{g}$ to the source code of the specification. 
Let $P=(C_{m}, C_{n}, c_{g})$ be an architectural specification and 
$G=(V_{com}, V_{con}, Com, Con, Int)$ be the AIFG of $P$. 
By using the concepts of a reduced component, connector, and configuration 
introduced in Section \ref{sec:sslicing}, a slice 
$S_{p} = (C_{m}', C_{n}', c_{g}')$ of an architectural specification 
$P$ can be constructed in the following steps:
 
\begin{figure*}[t]
  \renewcommand{\arraystretch}{0.85}
{\sf
\begin{center}
{\scriptsize
\begin{tabular}{l}
     {\bf Configuration} GasStation\\
     \qquad{\bf Component} Customer\\
     \qqqquad {\bf Port} Pay = \underline{pay!x} $\rightarrow$ Pay\\
     \qqqquad $\Box$$\Box$$\Box$$\Box$$\Box$$\Box$$\Box$$\Box$$\Box$$\Box$$\Box$$\Box$$\Box$$\Box$$\Box$$\Box$$\Box$$\Box$$\Box$$\Box$$\Box$$\Box$\\
     \qqqquad {\bf Computation} = \underline{Pay.pay!x} $\rightarrow$ \underline{Gas.take} $\rightarrow$ Gas.pump$?$x $\rightarrow$ Computation\\

     \qquad{\bf Component} Cashier\\
     \qqqquad {\bf Port} Customer1 = pay$?$x $\rightarrow$ Customer1\\
     \qqqquad {\bf Port} Customer2 = pay$?$x $\rightarrow$ Customer2\\
     \qqqquad {\bf Port} Topump = \underline{pump!x} $\rightarrow$ Topump\\
     \qqqquad {\bf Computation} = Customer1.pay$?$x $\rightarrow$ \underline{Topump.pump!x} $\rightarrow$ 
Computation\\
     \qqqqqquad [] Customer2.pay$?$x $\rightarrow$ \underline{Topump.pump!x} $\rightarrow$ Computation\\

     \qquad $\Box$$\Box$$\Box$$\Box$$\Box$$\Box$$\Box$$\Box$$\Box$$\Box$$\Box$$\Box$$\Box$$\Box$$\Box$$\Box$$\Box$\\
     \qqqquad $\Box$$\Box$$\Box$$\Box$$\Box$$\Box$$\Box$$\Box$$\Box$$\Box$$\Box$$\Box$$\Box$$\Box$$\Box$$\Box$$\Box$$\Box$$\Box$$\Box$$\Box$$\Box$\\
     \qqqquad $\Box$$\Box$$\Box$$\Box$$\Box$$\Box$$\Box$$\Box$$\Box$$\Box$$\Box$$\Box$$\Box$$\Box$$\Box$$\Box$$\Box$$\Box$$\Box$$\Box$$\Box$$\Box$\\
     \qqqquad $\Box$$\Box$$\Box$$\Box$$\Box$$\Box$$\Box$$\Box$$\Box$$\Box$$\Box$$\Box$$\Box$$\Box$$\Box$$\Box$$\Box$$\Box$$\Box$$\Box$$\Box$$\Box$$\Box$$\Box$$\Box$$\Box$\\
     \qqqquad $\Box$$\Box$$\Box$$\Box$$\Box$$\Box$$\Box$$\Box$$\Box$$\Box$$\Box$$\Box$$\Box$$\Box$$\Box$$\Box$$\Box$$\Box$$\Box$$\Box$$\Box$$\Box$$\Box$$\Box$$\Box$$\Box$$\Box$\\
     \qqqqqquad $\Box$$\Box$$\Box$$\Box$$\Box$$\Box$$\Box$$\Box$$\Box$$\Box$$\Box$$\Box$$\Box$$\Box$$\Box$$\Box$$\Box$$\Box$$\Box$$\Box$$\Box$$\Box$$\Box$$\Box$$\Box$$\Box$$\Box$\\
     \qqqqqquad $\Box$$\Box$$\Box$$\Box$$\Box$$\Box$$\Box$$\Box$$\Box$$\Box$$\Box$$\Box$$\Box$$\Box$$\Box$$\Box$$\Box$$\Box$$\Box$$\Box$$\Box$$\Box$$\Box$$\Box$$\Box$$\Box$$\Box$\\
     \qquad{\bf Connector} Customer\_Cashier\\
     \qqqquad {\bf Role} Givemoney = \underline{pay!x} $\rightarrow$ Givemoney\\
     \qqqquad {\bf Role} Getmoney = pay$?$x $\rightarrow$ Getmoney\\
     \qqqquad {\bf Glue} = Givemoney.pay$?$x $\rightarrow$ \underline{Getmoney.pay!x} $\rightarrow$ Glue\\
     \qquad $\Box$$\Box$$\Box$$\Box$$\Box$$\Box$$\Box$$\Box$$\Box$$\Box$$\Box$$\Box$$\Box$$\Box$$\Box$$\Box$$\Box$\\
     \qqqquad $\Box$$\Box$$\Box$$\Box$$\Box$$\Box$$\Box$$\Box$$\Box$$\Box$$\Box$$\Box$$\Box$$\Box$$\Box$$\Box$$\Box$$\Box$$\Box$$\Box$$\Box$$\Box$$\Box$$\Box$$\Box$$\Box$\\
     \qqqquad $\Box$$\Box$$\Box$$\Box$$\Box$$\Box$$\Box$$\Box$$\Box$$\Box$$\Box$$\Box$$\Box$$\Box$$\Box$$\Box$$\Box$$\Box$$\Box$$\Box$$\Box$$\Box$$\Box$$\Box$$\Box$$\Box$\\
     \qqqquad $\Box$$\Box$$\Box$$\Box$$\Box$$\Box$$\Box$$\Box$$\Box$$\Box$$\Box$$\Box$$\Box$$\Box$$\Box$$\Box$$\Box$$\Box$$\Box$$\Box$$\Box$$\Box$$\Box$$\Box$$\Box$$\Box$$\Box$$\Box$$\Box$$\Box$$\Box$$\Box$$\Box$$\Box$$\Box$$\Box$$\Box$\\
     \qquad $\Box$$\Box$$\Box$$\Box$$\Box$$\Box$$\Box$$\Box$$\Box$$\Box$$\Box$$\Box$$\Box$$\Box$$\Box$$\Box$$\Box$\\
     \qqqquad $\Box$$\Box$$\Box$$\Box$$\Box$$\Box$$\Box$$\Box$$\Box$$\Box$$\Box$$\Box$$\Box$$\Box$$\Box$$\Box$$\Box$$\Box$$\Box$$\Box$$\Box$\\
     \qqqquad $\Box$$\Box$$\Box$$\Box$$\Box$$\Box$$\Box$$\Box$$\Box$$\Box$$\Box$$\Box$$\Box$$\Box$$\Box$$\Box$$\Box$$\Box$$\Box$$\Box$$\Box$\\
     \qqqquad $\Box$$\Box$$\Box$$\Box$$\Box$$\Box$$\Box$$\Box$$\Box$$\Box$$\Box$$\Box$$\Box$$\Box$$\Box$$\Box$$\Box$$\Box$$\Box$$\Box$$\Box$$\Box$$\Box$$\Box$$\Box$$\Box$$\Box$\\
     \qquad{\bf Instances}\\
     \qqqquad Customer1: Customer\\
     \qqqquad Customer2: Customer\\
     \qqqquad cashier: Cashier\\
     \qqqquad $\Box$$\Box$$\Box$$\Box$$\Box$$\Box$$\Box$$\Box$\\
     \qqqquad Customer1\_cashier: Customer\_Cashier\\
     \qqqquad Customer2\_cashier: Customer\_Cashier\\
     \qqqquad $\Box$$\Box$$\Box$$\Box$$\Box$$\Box$$\Box$$\Box$$\Box$$\Box$$\Box$$\Box$$\Box$$\Box$$\Box$$\Box$$\Box$$\Box$$\Box$$\Box$$\Box$\\
     \qqqquad $\Box$$\Box$$\Box$$\Box$$\Box$$\Box$$\Box$$\Box$$\Box$$\Box$$\Box$$\Box$$\Box$$\Box$$\Box$$\Box$$\Box$$\Box$$\Box$$\Box$$\Box$\\
     \qqqquad $\Box$$\Box$$\Box$$\Box$$\Box$$\Box$$\Box$$\Box$$\Box$$\Box$$\Box$$\Box$$\Box$$\Box$$\Box$$\Box$$\Box$$\Box$\\
     \qquad{\bf Attachments}\\
     \qqqquad Customer1.Pay as Customer1\_cashier.Givemoney\\
     \qqqquad $\Box$$\Box$$\Box$$\Box$$\Box$$\Box$$\Box$$\Box$$\Box$$\Box$$\Box$$\Box$$\Box$$\Box$$\Box$$\Box$$\Box$$\Box$$\Box$$\Box$$\Box$$\Box$$\Box$$\Box$$\Box$$\Box$$\Box$\\
     \qqqquad Customer2.Pay as Customer2\_cashier.Givemoney\\
     \qqqquad $\Box$$\Box$$\Box$$\Box$$\Box$$\Box$$\Box$$\Box$$\Box$$\Box$$\Box$$\Box$$\Box$$\Box$$\Box$$\Box$$\Box$$\Box$$\Box$$\Box$$\Box$$\Box$$\Box$$\Box$$\Box$$\Box$$\Box$\\
     \qqqquad casier.Customer1 as Customer1\_cashier.Getmoney\\
     \qqqquad casier.Customer2 as Customer2\_cashier.Getmoney\\
     \qqqquad $\Box$$\Box$$\Box$$\Box$$\Box$$\Box$$\Box$$\Box$$\Box$$\Box$$\Box$$\Box$$\Box$$\Box$$\Box$$\Box$$\Box$$\Box$$\Box$$\Box$$\Box$$\Box$$\Box$$\Box$\\
     \qqqquad $\Box$$\Box$$\Box$$\Box$$\Box$$\Box$$\Box$$\Box$$\Box$$\Box$$\Box$$\Box$$\Box$$\Box$$\Box$$\Box$$\Box$$\Box$$\Box$$\Box$$\Box$$\Box$$\Box$$\Box$$\Box$$\Box$$\Box$\\
     \qqqquad $\Box$$\Box$$\Box$$\Box$$\Box$$\Box$$\Box$$\Box$$\Box$$\Box$$\Box$$\Box$$\Box$$\Box$$\Box$$\Box$$\Box$$\Box$$\Box$$\Box$$\Box$$\Box$$\Box$$\Box$\\
     \qqqquad $\Box$$\Box$$\Box$$\Box$$\Box$$\Box$$\Box$$\Box$$\Box$$\Box$$\Box$$\Box$$\Box$$\Box$$\Box$$\Box$$\Box$$\Box$$\Box$$\Box$$\Box$$\Box$$\Box$$\Box$\\
     \qquad{\bf End} GasStation.\\
    \\
\end{tabular}
}
\end{center}
}

  \caption{\label{fig:wright.slice} A backward slice of the architectural 
specification in Figure {\protect\ref{fig:wright}}.}

\end{figure*}

\begin{itemize}

\item[1.] Constructing a reduced component $c_{m}'$ from a component 
           $c_{m}$ by removing all ports such that their corresponding 
           port vertices in $G$ have not been included in $S_{g}$ and 
           unnecessary elements in the computation from $c_{m}$. 
           The reduced components $C_{m}'$ in $S_{p}$ have the same
           relative order as the components $C_{m}$ in $P$.

\item[2.] Constructing a reduced connector $c_{n}'$ from a connector 
           $c_{n}$ by removing all roles such that their corresponding 
           role vertices in $G$ have not been included in $S_{g}$ and 
           unnecessary elements in the glue from $c_{n}$. 
           The reduced connectors $C_{n}'$ in $S_{p}$ have the same 
           relative order as their corresponding connectors in $P$.

\item[3.] Constructing the reduced configuration $c_{g}'$ from the 
           configuration $c_{g}$ by the following steps:

\begin{itemize}
	\item Removing all component and connector instances from $c_{g}$ 
              that are not included in $C_{m}'$ and $C_{n}'$. 

	\item Removing all attachments from $c_{g}$ such that there exists 
              no two vertices $v_{1}$ and $v_{2}$ where 
              $v_{1}, v_{2} \in S_{g}$ and \verb+v1 as v2+ represents 
              an attachment.
              
	\item The instances and attachments in the reduced configuration 
              in $S_{p}$ have the same relative order as their corresponding 
              instances and attachments in $P$.
\end{itemize} 

\end{itemize}

Figure \ref{fig:wright.slice} shows a backward slice of the 
W{\sc right} specification in Figure \ref{fig:wright} with respect to 
the slicing criterion (\verb+cashier+, E) such that 
E=\{\verb+Customer1+, \verb+Customer2+, \verb+Topump+\}
is a set of ports of component \verb+cashier+.
The small rectangles represent the parts of specification that have been
removed, i.e., sliced away from the original specification. 
The slice is obtained from a slice over the AIFG in Figure
\ref{fig:graph.slice} according to the mapping process described above. 
Figure \ref{fig:sys-slice} shows the architectural representation of
the slice in Figure \ref{fig:wright.slice}.

\begin{figure}[t]
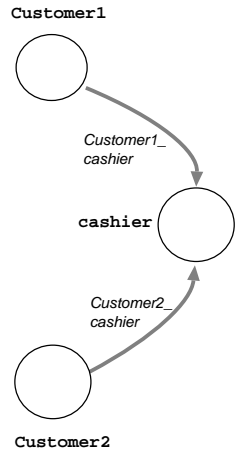

  \begin{center}
     \epsfile{file=fig5.eps,scale=0.85}
  \caption{\label{fig:sys-slice} The architectural representation of
the slice in Figure {\protect\ref{fig:wright.slice}}.}
  \end{center}
\end{figure}

\vspace{3mm}
\section{Related Work}
\label{sec:work}
\vspace{3mm}

\vspace{2mm}
\subsection{Software Architecture Dependence Analysis} 
\vspace{2mm}

Perhaps, the most similar work with ours is that 
presented by Stafford, Richardson and Wolf \cite{Stafford97}, who 
introduced a software architecture dependence analysis technique, 
called {\it chaining} to support software architecture development such 
as debugging and testing. In chaining, links represent the dependence 
relationships that exist in an architectural specification. Links connect 
elements of the specification that are directly related, producing a chain of 
dependences similar to a slice in traditional slicing that can be followed 
during analysis. Although their consideration is similar to ours, there 
are still some differences between their work and ours. First, 
the slicing criterions are different. While Stafford, Richardson, 
and Wolf define a slicing criterion of an architectural specification 
as a set of ports of a component, we defined a slicing criterion as either 
a set of ports of a component or a set of roles of a connector of an 
architectural specification. This is because that in addition to modifying 
a component, in some cases, a maintainer may also want to modify a connector. 
Second, the types of architectural slices are different. 
Stafford, Richardson, and Wolf compute an architectural slice that 
includes only a set of components of an architectural specification, and 
therefore, it seems that their slices fail to capture the information 
concerning interactions between these components. In contrast, we compute 
an architectural slice that includes not only a set of components but also 
connectors (interactions between these components). Moreover, since 
our architectural slice is a reduced architectural specification 
of the original one and can also preserve the partial semantics of the 
original architectural specification, our slice is particularly useful 
in software architecture reuse.

\vspace{2mm}
\subsection{Class Slicing for C++}
\vspace{2mm}

Tip {\it et.al} \cite{Tip96} introduced an algorithm for slicing 
class hierarchies in C++ programs. Given a C++ class hierarchy 
(a collection of C++ class and inheritance relations among them) 
and a program that uses the hierarchy, the algorithm eliminates 
from the hierarchy those data members, member of
functions, classes, and inheritance relations that are unnecessary for
ensuring that the semantics of the program is maintained. 
The class slicing has the benefit of allowing unused components of
classes to be eliminated in applications that do not use those
components. In this aspect, our work is strongly inspired by their work in 
the sense that we also want to use architectural slicing to remove 
unused components at the architectural level of software systems 
to narrow the domain on which reasoning about bugs is performed 
during the debugging at the architectural level. 

\vspace{2mm}
\subsection{Generalized Slicing}
\vspace{2mm}

Another work beyond traditional slicing is presented by Sloane
and Holdsworth \cite{Sloane96}. They observed that two assumptions
implicit in the definition of a traditional slice for programs written
in imperative programming languages: (1) that variables
and statements are concepts of the programming language in which
program is written, and (2) that slices consist only of statements. For
a language that does not have variables and statements, for example, a
compiler specification language, traditional slicing does not make
sense. To solve this problem, they introduced the generalized slicing 
as an extension of the traditional slicing by replacing variables
with arbitrary named program entities and statements with arbitrary
program constructs. This allows them to perform the slicing of 
non-imperative programs. Our work has a similar goal with theirs, but 
focuses specially on software architectures.
 
\vspace{3mm}
\section{Concluding Remarks}
\label{sec:final}
\vspace{3mm}

\noindent
We introduced a new form of slicing, named 
{\it architectural slicing} to aid architectural understanding and reuse. 
In contrast to the traditional
slicing, architectural slicing is designed to operate on the architectural 
specification of a software system, rather than the source code of a
program. Architectural slicing provides knowledge about the high-level 
structure of a software system, rather than the low-level
implementation details of a program. In order to compute an
architectural slice, we presented the {\it architecture information 
flow graph} to explicitly represent information flows in a formal 
architectural specification. Based on the graph, we gave a 
two-phase algorithm to compute an architectural slice.

While our initial exploration used W{\sc right} as the architecture 
description language, the concept and approach of architectural slicing
are language-independent. However, the implementation of 
an architectural slicing tool may differ from one architecture description
language to another because each language has its own structure and
syntax which must be handled carefully. 

In architectural description languages, in addition to provide both
a conceptual framework and a concrete syntax for characterizing
software architectures, they also provide tools for parsing, 
displaying, compiling, analyzing, or simulating architectural 
specifications written in their associated language. However, 
existing language environments provide no tools to support 
architectural understanding, maintenance, testing, and reuse 
from an engineering viewpoint. We believe that some static analysis 
tools such as an architectural slicing tool introduced in this paper and an 
architectural dependence analysis tool \cite{Stafford97,Zhao97a} should be 
provided by any ADL as an essential means to support these development
activities.

As future work, we would like to extend our approach presented here 
to handle other constructs in W{\sc right} language such as {\it styles} 
which were not considered here, and 
also to extend our approach to handle the slicing problem for 
other architecture description languages such as Rapide, ACME, and UniCon. 
To demonstrate the usefulness of our slicing approach, 
we are implementing a slicer for W{\sc right} architectural descriptions to 
support architectural level understanding and reuse. 
The next step for us is to perform some experiments to evaluate the 
usefulness of architectural slicing in practical development of 
software architectures. 
}

\vspace{3mm}
\noindent
{\large\bf  Acknowledgements}
\vspace{3mm}
 
The author would like to thank the anonymous referees for their valuable 
suggestions and comments on earlier drafts of the paper.

\vspace{3mm}

\renewcommand{\baselinestretch}{.9}


\end{document}